\documentclass[prl, twocolumn, reprint, nofootinbib]{revtex4-2}

\usepackage{graphicx}
\usepackage{epsfig}
\usepackage{epstopdf}
\usepackage{amsmath}
\usepackage{amsfonts}

\newcommand{\be}{\begin{equation}}
\newcommand{\ee}{\end{equation}}
\newcommand{\bea}{\begin{eqnarray}}
\newcommand{\eea}{\end{eqnarray}}

\begin{document}


\title{Hunting Inflaton at FASER}

\author{Nobuchika Okada}
\email{okadan@ua.edu}
\affiliation{
Department of Physics and Astronomy,
University of Alabama,
Tuscaloosa, AL 35487, USA
}
\author{Digesh Raut}
\email{draut@udel.edu}
\affiliation{
Bartol Research Institute,
Department of Physics and Astronomy,
University of Delaware, Newark, DE 19716, USA
}


\begin{abstract}
We consider a non-minimal quartic inflation scenario in the minimal U(1)$_X$ extension of the Standard Model (SM) 
   with the classical conformal invariance, where the inflaton is identified with the U(1)$_X$ Higgs field ($\phi$).  
By virtue of the classically conformal invariance and the radiative U(1)$_X$ symmetry breaking 
   via the Coleman-Weinberg mechanism, 
   the inflationary predictions (in particular, the tensor-to-scaler ratio ($r$)), the U(1)$_X$ coupling ($g_X$) 
   and the U(1)$_X$ gauge boson mass ($m_{Z^\prime}$), are all determined by only two free parameters, 
   the inflaton mass ($m_\phi$) and its mixing angle ($\theta$) with the SM Higgs field.   
FASER can search for a long-lived scalar, which is the inflaton in our scenario, for the parameter ranges of $0.1 \lesssim m_\phi[{\rm GeV}] \lesssim 4$
   and $10^{-5} \lesssim \theta \lesssim 10^{-3}$. 
Therefore, once such a scalar is discovered at FASER, both $m_\phi$ and $\theta$ are fixed, leading to the prediction for $r$, $g_X$ and $m_{Z^\prime}$ in our model. 
These predictions can be tested by the future cosmological observation and LHC search for $Z^\prime$ boson resonance. 
\end{abstract}

\maketitle


Very recently, the ForwArd Search ExpeRiment (FASER) \cite{Ariga:2019ufm} 
  has been approved to search for light, weakly interacting, electrically neutral long-lived particles 
  at the Large Hadron Collider (LHC). 
Such long-lived particles are included in a variety of new physics models beyond the Standard Model (SM). 
In the experiment, a detector will be located along the beam trajectory 480 meters downstream 
  from the interaction point within the ATLAS detector at the LHC. 
This setup is specialized to search for light, long-lived particles with the following advantages: 
(i) the High-Luminosity upgrade of the LHC (HL-LHC) can produce a huge number of hadrons 
in the forward region, which could decay into light long-lived particles. 
Even if such a decay process is extremely rare, the huge number of produced hadrons 
  provides us with a sizable number of events for the long-lived particle production;  
(ii) such light particles are highly boosted in the beam direction and mostly produced in the forward region;  
(iii) Because of very weak interactions, such particles can have a decay length of $\mathcal{O}$(100 m).  
The displaced vertex signature from such long-lived particles is almost free from the SM backgrounds. 
In Refs.~\cite{Feng:2017vli, Ariga:2018uku}, 
   the authors have explored a possibility of detecting a SM singlet scalar ($\phi$) 
   at FASER and other proposed experiments for the displaced vertex search. 
The singlet scalar only couples with the SM particles through its mixing with the SM Higgs boson. 
Hence, the production rate and the lifetime of the particle $\phi$ are controlled by only two parameters, 
   its mass ($m_\phi$) and mixing angle ($\theta$) with the SM Higgs field. 
Impressively, these experiments are capable of probing extremely small mixing angles, $10^{-7} \lesssim \theta \lesssim 10^{-3}$, 
   for $0.1\lesssim m_\phi [{\rm GeV}] \lesssim10$ \cite{Feng:2017vli, Ariga:2018uku}.

Authors of the pioneering work \cite{Bezrukov:2009yw} have pointed out that the long-lived light scalar can be identified with a light inflaton in their inflation model. 
Once observed, its mass and mixing angle with the SM Higgs field can be measured. 
In their model, the measured mass and mixing angle correspond to the inflationary predictions. Although the quartic inflation scenario discussed in Ref.~\cite{Bezrukov:2009yw} is excluded by Planck 2018 results \cite{Akrami:2018odb}, it can be made consistent with the Planck 2018 by including non-minimal gravitational coupling \cite{Bezrukov:2013fca}.  

In this letter, we consider the non-minimal quartic inflation in a classically conformal U(1)$_X$ extended SM, which the authors of the present paper have proposed with their collaborators \cite{Oda:2017zul} (see also Ref.~\cite{Marzola:2016xgb}). 
By imposing the conformal invariance at the classical level on the minimal U(1)$_X$ extended SM \cite{Oda:2015gna}, 
  all the mass terms in the Higgs potential is forbidden. 
As a result, the U(1)$_X$ gauge symmetry is radiatively broken by the Coleman-Weinberg (CW) mechanism \cite{CW}, 
  which subsequently drives the electroweak symmetry breaking through a mixing quartic coupling 
  between the U(1)$_X$ Higgs and the SM Higgs fields \cite{Iso:2009ss}. 
As first pointed out in Ref.~\cite{Bardeen:1995kv}, the classical conformal invariance could be a clue
  to solve the gauge hierarchy problem of the SM.  
In our paper \cite{Oda:2017zul}, we have identified the U(1)$_X$ Higgs field with a non-minimal gravitational coupling as inflaton. 
Because of the classically conformal invariance, this scenario not only leads to the inflationary predictions consistent
  with the Planck 2018 results \cite{Akrami:2018odb}, but also provides a direct connection 
  between the inflationary predictions and the LHC search for the U(1)$_X$ gauge boson ($Z^\prime$) resonance. 
In our previous work \cite{Oda:2017zul}, 
the scalar sector was not analyzed in detail from the viewpoint of collider phenomenology. 
The main purpose of this letter is to point out that if the inflaton mass and its mixing angle with the SM Higgs field 
  lie in a suitable range, the inflaton can be searched by FASER 
  with a direct connection to the inflationary predictions.\footnote{See Ref.~\cite{Daido:2017wwb} for a similar work for an axion-like particle as the inflaton.}
Therefore, three independent experiments, namely, 
  the inflaton search at FASER, the $Z^\prime$ boson resonance search at the HL-LHC 
  and the precision measurement of the inflationary predictions, 
  are complementary to test our inflation scenario.

\begin{table}[t]
\begin{center}
\begin{tabular}{|c|ccc|c|}
\hline
      &  SU(3)$_c$  & SU(2)$_L$ & U(1)$_Y$ & U(1)$_X$  \\ 
\hline
$q^{i}_{L}$ & {\bf 3 }    &  {\bf 2}         & $ 1/6$       & $(1/6) x_{H} + (1/3)$   \\
$u^{i}_{R}$ & {\bf 3 }    &  {\bf 1}         & $ 2/3$       & $(2/3) x_{H} + (1/3)$   \\
$d^{i}_{R}$ & {\bf 3 }    &  {\bf 1}         & $-1/3$       & $(-1/3) x_{H} + (1/3)$  \\
\hline
$\ell^{i}_{L}$ & {\bf 1 }    &  {\bf 2}         & $-1/2$       & $(-1/2) x_{H} -1$    \\
$e^{i}_{R}$    & {\bf 1 }    &  {\bf 1}         & $-1$                   & $- x_{H} -1$   \\
\hline
$H$            & {\bf 1 }    &  {\bf 2}         & $- 1/2$       & $(-1/2) x_{H}$   \\  
\hline
$N^{i}_{R}$    & {\bf 1 }    &  {\bf 1}         &$0$                    & $-1$     \\
$\Phi$            & {\bf 1 }       &  {\bf 1}       &$ 0$                  & $ 2$  \\ 
\hline
\end{tabular}
\end{center}
\caption{
The particle content of the minimal U(1)$_X$ model. 
$i = 1,2,3$ is the generation index. }
\label{table2}
\end{table}

{\bf Classically conformal U(1)$_X$ model}: We first define our model with the particle content listed in Table~\ref{table2}, 
  where the U(1)$_X$ charge of a particle is defined as a linear combination of its SM hypercharge 
  and its ${B-L}$ (Baryon minus Lepton) number. 
The U(1)$_X$ charges are determined by a real parameter, $x_H$, and 
   the well-known minimal U(1)$_{B-L}$ model \cite{mBL} is realized as the limit of $x_H \to 0$. 
In the presence of the three right-hand neutrinos (RHNs), $N_R^{1,2,3}$, 
  this model is free from all the gauge and the mixed gauge-gravitational anomalies. 
Once the U(1)$_{X}$ Higgs field ($\Phi$) develops a vacuum expectation value (VEV), 
  $\langle \Phi \rangle = v_X/\sqrt{2}$, the U(1)$_{X}$ gauge symmetry is broken 
  and the $Z^\prime$ boson becomes massive, $m_{Z^\prime}  =  2 g_{X} v_{X}$, 
  where $g_X$ is the U(1)$_X$ gauge coupling.  

The Yukawa sector of the SM is extended to include  
\bea
\mathcal{L}_{Y} \supset  - \sum_{i, j=1}^{3} Y^{ij}_{D} \overline{\ell^i_{L}} H N_R^j 
          -\frac{1}{2} \sum_{k=1}^{3} Y_M^k \Phi \overline{N_R^{k~C}} N_R^k , 
\label{Lag1} 
\eea
where $Y_D$ ($Y_M$) is a Dirac (Majorana) type Yukawa coupling. 
Without a loss of generality, we chose the Majorana Yukawa couplings to be flavor diagonal.  
The Majorana masses for the RHNs are generated by the U(1)$_{X}$ gauge symmetry breaking.  
For simplicity, we fix $Y_M^{1,2,3} = Y_M$ and thus RHNs have a degenerate mass spectrum, $ m_N = Y_M v_{X}/{\sqrt 2}$.
After the electroweak symmetry breaking, the light neutrino masses are generated via the type-I seesaw mechanism \cite{Seesaw}.

Imposing the classical conformal invariance, the Higgs potential of our model is given by
\bea  
V = \lambda_H \!\left(  H^{\dagger}H  \right)^2
+ \lambda_{\Phi}\! \left(  \Phi^{\dagger} \Phi   \right)^2
- \lambda_{\rm mix} \!
\left(  H^{\dagger}H   \right) \!
\left(  \Phi^{\dagger} \Phi  \right), \qquad 
\label{Higgs_Potential}
\eea
where we set $\lambda_{H,\Phi,{\rm mix}} > 0$.  
Assuming $\lambda_{\rm mix} \ll1$ (this will be justified later), 
  we can separately analyze the Higgs potential for $\Phi$ and $H$ \footnote{In our analysis, only a non-zero $\lambda_{\rm mix}\ll 1$ at $v_X$ is important. Its contribution to the inflaton potential at high energies through the RG evolution is negligibly small \cite{Oda:2015gna}.}. 
The CW potential for the Higgs field $\Phi$ at the 1-loop level is given by \cite{CW}
\begin{eqnarray}
   V(\phi) =  \frac{\lambda_\Phi}{4} \phi^4 
     + \frac{\beta_\Phi}{8} \phi^4 \left(  \ln \left[ \frac{\phi^2}{v_{X}^2} \right] - \frac{25}{6} \right), 
\label{eq:CW_potential} 
\end{eqnarray}
where $\phi = \sqrt{2}\Re[\Phi]$, $v_{\rm X}$ is chosen as a renormalization scale, 
 and the coefficient of the 1-loop corrections is approximately given by
\begin{eqnarray}
 16 \pi^2 \beta_\Phi \simeq 96 g_{X}^4  - 3 Y_M^4.   
\label{eq:beta}
\end{eqnarray} 
The stationary condition, $\left. dV/d\phi\right|_{\phi=v_{X}} = 0$, leads to  
\begin{eqnarray}
{\overline {\lambda_\Phi}} = \frac{11}{6} {\overline {\beta_\Phi}},
\label{eq:stationary}
\end{eqnarray}  
where the {\it barred} quantities are evaluated at $ \langle \phi \rangle = v_X$. 
The mass of $\phi$ is given by 
\begin{eqnarray}
  m_\phi^2 &=& \left. \frac{d^2 V}{d\phi^2}\right|_{\phi=v_{X}}    
={\overline {\beta_\Phi}} v_{X}^2  \nonumber \\ 
&=& 
    \frac{6}{\pi} {\overline{\alpha_{X}}}m_{Z^\prime}^2 
     \left( 1-2 \left( \frac{m_N}{m_{Z^\prime}}\right)^4   \right), 
\label{eq:mass_phi}
\end{eqnarray}
where $\alpha_{X} = g_{X}^2/(4 \pi)$. 
The condition for the stability of $U(1)_X$ vacuum, $m_\phi^2 >0$, requires $m_{Z^\prime} >  2^{1/4} m_N$.

The U(1)$_X$ gauge symmetry breaking by $\langle \Phi \rangle = v_X/\sqrt{2}$ induces 
  a negative mass squared for the SM Higgs doublet ($-\lambda_{\rm mix} |\langle \Phi \rangle|^2$) 
  in Eq.~(\ref{Higgs_Potential}) and triggers the electroweak symmetry breaking \cite{Iso:2009ss}. 
The SM(-like) Higgs boson mass ($m_h = 125$ GeV) is described as  
\begin{equation}
  m_h^2  = \lambda_{\rm mix} v_{X}^2 = 2 \lambda_H v_h^2, 
\label{eq:lambdamix}
\end{equation} 
where $v_h = 246$ GeV is the Higgs doublet VEV. 
 From this formula, we can justify our assumption of $\lambda_{\rm mix} \ll 1$
   by considering the LEP constraint on $v_X \gtrsim 10$ TeV 
   \cite{LEP:2003aa, Carena:2004xs, Schael:2013ita, Heeck:2014zfa}.

The mass matrix for the Higgs bosons, $\phi$ and $h$, is given by
\begin{eqnarray}
{\cal L}  \supset -
\frac{1}{2}
\begin{bmatrix}
h & \phi
\end{bmatrix}
\begin{bmatrix} 
m_{h}^2 &  \lambda_{\rm mix} v_{X} v_{h} \\ 
 \lambda_{\rm mix} v_{X} v_{h} & m_{\phi}^2
\end{bmatrix} 
\begin{bmatrix} 
h \\ \phi 
\end{bmatrix}.  
\label{eq: massmatrix}
\end{eqnarray} 
We diagonalize the mass matrix by 
\begin{eqnarray}
\begin{bmatrix} 
h \\ \phi 
\end{bmatrix}
=
\begin{bmatrix} 
\cos\theta &   \sin\theta \\ 
-\sin\theta & \cos\theta  
\end{bmatrix} 
\begin{bmatrix} 
{\tilde h} \\ {\tilde \phi} 
\end{bmatrix}  ,
\label{eq: eigenstate}
\end{eqnarray} 
where ${\tilde h}$ and ${\tilde \phi}$ are the mass eigenstates, and 
   the mixing angle $\theta$ is determined by 
\bea
2 v_{X} v_{h}  \lambda_{\rm mix}= ( m_h^2 -m_\phi^2) \tan2\theta.  
\label{eq: mixings} 
\eea  
Since we are interested in the case with $m_\phi^2  \ll m_h^2$ and $\lambda_{\rm mix} \ll1$, 
  we find
\bea
\theta \simeq \frac{v_h}{v_X} = \frac{\sqrt{16 \pi \overline{\alpha_X}} v_h}{m_{Z^\prime}}  \ll 1. 
\label{eq:theta}
\eea
The mass eigenvalues are given by
\bea
&&m_{\tilde{\phi}}^2 = m_{\phi}^2  + \left(m_\phi^2  - m_h^2 \right) \frac{\sin^2\theta}{1-2 \sin^2\theta} 
  \simeq m_{\phi}^2   - m_h^2 \theta^2,   \nonumber\\
&&m_{\tilde{h}}^2 = m_h^2 - \left(m_\phi^2  - m_h^2 \right) \frac{\sin^2\theta}{1-2 \sin^2\theta} \simeq m_h^2. 
\label{eq: masses} 
\eea  
For the parameter region which will be searched by FASER, 
  we find $m_{\tilde{\phi}, {\tilde{h}}}  \simeq m_{\phi, h}$ and $\tilde{\phi}, \tilde{h} \simeq \phi, h$.  
For notational simplicity, we will refer to the mass eigenstates without using {\it tilde} in the rest of this letter. 
Note that for a fixed value of $m_N/m_{Z^\prime}$, 
   the inflaton mass ($m_\phi$) and its mixing angle with the Higgs field ($\theta$)
   are uniquely determined by ${\overline {\alpha_X}}$ and $m_{Z^\prime}$ 
   with Eqs.~(\ref{eq:mass_phi}) and (\ref{eq:theta}).

{\bf non-minimal quartic inflation}: We here give a brief review on non-minimal quartic inflation
with the action in the Jordan frame:  
\begin{eqnarray}
 {\cal S}_J &=& \int d^4 x \sqrt{-g} 
   \left[-\frac{1}{2} f(\phi)  {\cal R}+ \frac{1}{2} g^{\mu \nu} \left(\partial_\mu \phi \right) \left(\partial_\nu \phi \right) 
\right.
\nonumber \\ 
&&\left. 
 - V_J (\phi) \right]  , 
\label{S_J}
\end{eqnarray}
where $\phi$ is a real scalar field (inflaton), $f(\phi) = (1+ \xi \phi^2)$ with a real parameter $\xi > 0$, 
  $V_J (\phi)= \lambda \phi^4/4$ is the inflaton quartic potential, 
  and the reduced Planck mass of $M_P=2.44 \times 10^{18}$ GeV is set to be 1 (Planck unit). 
Using the transformation of $f(\phi) g_{\mu \nu} =  g_{E {\mu \nu}}$, 
  the action in the Einstein frame is described as 
\begin{eqnarray}
S_E &=& \int d^4 x \sqrt{-g_E}\left[-\frac{1}{2}  {\cal R}_E +  \frac{1}{2} g_E^{\mu \nu} \left(\partial_\mu \sigma \right) \left(\partial_\nu \sigma \right)
\right.
\nonumber \\ 
&&\left. 
   -V_E(\phi(\sigma)) \right], 
\label{S_E}   
\end{eqnarray}
where $V_E(\phi(\sigma)) = V_J (\phi)/ f(\phi)^2$, and $\sigma$ is a canonically normalized scalar field 
  (inflaton in the Einstein frame) which is related to the original field $\phi$ by
\begin{eqnarray}
\left(\frac{d\sigma}{d\phi}\right)^{2} = \frac{1+ \xi (6 \xi +1) \phi^2} {\left( 1 + \xi \phi^2 \right)^2}. 
\label{eq:sigphi}
\end{eqnarray}
Using Eq.~(\ref{eq:sigphi}), we can express the slow-roll inflation parameters in the Einstein frame as 
\begin{eqnarray}
 \epsilon(\phi) &=& \frac{1}{2} \left(\frac{V_E^\prime}{V_E \; \sigma^\prime}\right)^2,   \nonumber \\
 \eta(\phi) &=& \frac{V_E^{\prime \prime}}{V_E \; (\sigma^\prime)^2}- \frac{V_E^\prime \; \sigma^{\prime \prime}}{V_E \; (\sigma^\prime)^3} ,   \nonumber \\
 \zeta (\phi) &=&  \left(\frac{V_E^\prime}{V_E \; \sigma^\prime}\right) 
 \left( \frac{V_E'''}{V_E \; (\sigma^\prime)^3}
-3 \frac{V_E'' \; \sigma''}{V_E \; (\sigma^\prime)^4} 
\right.
\nonumber \\ 
&&\left. 
+ 3 \frac{V_E^\prime \; (\sigma^{\prime \prime})^2}{V_E \; (\sigma^\prime)^5} 
- \frac{V_E^\prime \; \sigma'''}{V_E \; (\sigma')^4} \right)  , 
\end{eqnarray}
where a {\it prime} denotes a derivative with respect to $\phi$.  
The slow-roll inflation takes place when $\epsilon,  |\eta|, \zeta \ll 1$. 
The amplitude of the curvature perturbation, 
\begin{equation} 
  \Delta_{\cal R}^2 = \left. \frac{V_E (\phi)}{24 \pi^2 \epsilon (\phi) } \right|_{k_0},
\end{equation}
should satisfy $\Delta_\mathcal{R}^2= 2.099 \times10^{-9}$ from the Planck 2018 result \cite{Akrami:2018odb} 
  for the pivot scale $k_0 = 0.05$ Mpc$^{-1}$.
The number of e-folds is evaluated by
\begin{eqnarray}
  N_0 = \frac{1}{\sqrt{2}} \int_{\phi_{\rm e}}^{\phi_0}
  d \phi  \frac{\sigma^\prime}{\sqrt{\epsilon(\phi)}}
\end{eqnarray} 
where $\phi_0$ is the inflaton value at the horizon exit of the scale corresponding to $k_0$,  
   and $\phi_e$ is the inflaton value at the end of inflation, which is defined by $\epsilon(\phi_e)=1$.
In our analysis, we fix $N_0=50$ to solve the horizon and flatness problems.

The inflationary predictions for the scalar spectral index ($n_{s}$), the tensor-to-scalar ratio ($r$), 
  and the running of the spectral index ($\alpha=\frac{d n_{s}}{d \ln k}$), are given by
\begin{eqnarray}
n_s = 1-6\epsilon+2\eta, \; 
r = 16 \epsilon,  \;  
\alpha \!=16 \epsilon \eta - 24 \epsilon^2 - 2 \zeta, 
\end{eqnarray} 
which are evaluated at $\phi=\phi_0$. 
Using $\Delta_\mathcal{R}^2= 2.099 \times10^{-9}$ and $N_0 = 50$,  
  the inflationary predictions, $\lambda$, $\phi_0$ and $\phi_e$ 
  are determined as a function of the non-minimal gravitational coupling $\xi$. 
Based on unitarity arguments \cite{Unitarity}, we only consider $\xi < 10$. 
Our results are summarized in Table~\ref{Tab:1}. 

\begin{table}[t!]
\begin{center}
\begin{tabular}{|c||cc|ccc|c|}
\hline 
 $\xi $ &  $\phi_0/M_p$ & $\phi_e/M_p $ & $n_s$   &  $r$  &  $\alpha (10^{-4})$  & $\lambda$\\ 
\hline
$0$             & $20.2$  & $2.83$    &  $0.941$ &  $ 0.314$  & $-11.5 $    &  $ 2.45  \times 10^{-13}  $    \\
$0.00527$  & $20.05$  & $2.77$    & $0.954$  &  $ 0.1$     & $-9.74$      &  $ 7.83  \times 10^{-13}  $    \\
$0.00978$  & $19.84$  & $2.73$    & $0.957$  &  $ 0.064$     & $-9.06$      &  $ 1.26 \times 10^{-13}  $  \\
$0.119$    & $15.75$    & $2.06$    & $0.961$  &  $ 0.010$   & $-7.70$        &  $ 1.97  \times 10^{-12}  $    \\
$1$             & $7.82$      & $1.00$       & $0.962$  &  $0.0049$    & $-7.51$ &  $ 6.56  \times 10^{-10}$  \\
$10$           & $2.65$      & $0.337$   & $0.962$  &  $0.0043$  & $-7.49$     &  $ 5.70  \times 10^{-8} $    \\
\hline
\end{tabular}
\end{center}
\caption{ 
Inflationary predictions for various $\xi$ values and $N_0=50$. 
The region $\xi < 0.00642$ ($r > 0.064$) is excluded by the Planck 2018 result. 
} 
\label{Tab:1}
\end{table}

{\bf Non-minimal U(1)$_X$ Higgs inflaton}: By introducing the non-minimal graviational coupling 
  of $- \xi  \left(  \Phi^\dagger \Phi \right)  {\cal R}$, 
  we identify the U(1)$_X$ Higgs field with the inflaton field in Eq.~(\ref{S_J}). 
Since  $\phi \gg v_X$ during inflation, we approximate the Higgs potential by its quartic potential 
  in the following inflationary analysis.

For the inflation analysis, we employ the renormalization group (RG) improved effective potential of the form \cite{Sher:1988mj},
\bea 
  V(\phi) = \frac{1}{4} \lambda_\Phi(\phi)  \phi^4 ,
\eea 
where $ \lambda_\Phi(\phi)$ is the solution to the following RG equations at the 1-loop level:
\bea 
  \frac{d \lambda_\Phi}{d \ln \phi} &=& \beta_{\lambda} \simeq 
   96 \alpha_X^2 - 3 \alpha_Y^2,  \nonumber \\
  \frac{d \alpha_X}{d \ln \phi } &=& \beta_g = \frac{72 + 64 x_H + 41 x_H^2}{12 \pi}  \alpha_X^2,  \nonumber \\
\frac{d \alpha_Y}{d \ln \phi} &=& \beta_Y = 
    \frac{1}{2 \pi} \alpha_Y 
   \left( \frac{5}{2} \alpha_Y - 6  \alpha_X \right) .
\label{eq:ApproxRGE}  
\eea
Here, $\alpha_Y=Y_M^2/(4 \pi)$ and we have identified $\phi$ with the renormalization scale along the inflation trajectory.

Since $\lambda_\Phi \ll 1$, the stationary condition in  Eq.~(\ref{eq:stationary}) implies that $g_X$, $Y_M \ll 1$. 
Hence, the RG evolutions of $\alpha_X$ and $\alpha_Y$ can be approximated as 
\bea 
  \alpha_{X, Y}(\phi) \simeq  \overline{\alpha_{X, Y}} + \overline{\beta_{g, Y}} \ln \left[\frac{\phi}{v_X} \right], 
 \label{eq:alphasol}
\eea 
and accordingly, 
\bea
 \beta_\lambda (\phi)
\simeq \overline{\beta_\lambda} + 2 \left( 96 \; \overline{\alpha_X} \; \overline{\beta_g} -3  \; \overline{\alpha_Y} \; \overline{\beta_Y} 
 \right) \ln \left[\frac{\phi}{v_X} \right].   
\eea
We now approximate the evolution of the quartic coupling by
\bea 
 \lambda_\Phi(\phi) &\simeq &
   \left( \frac{11}{6} + \ln \left[\frac{\phi}{v_X} \right] \right) \overline{\beta_\lambda} 
\nonumber \\
  &+& \left( 96 \; \overline{\alpha_X} \; \overline{\beta_g} -3  \; \overline{\alpha_Y} \; \overline{\beta_Y} 
 \right) \left( \ln \left[\frac{\phi}{v_X} \right] \right)^2  . 
\label{RGEsol}
\eea

In the following analysis, we fix $m_N = m_{Z^\prime}/3$ (or equivalentely, $\overline{\alpha_Y} = 8  \overline{\alpha_X}/9$)
  to satisfy the vacuum stability condition\footnote{As long as $m_N^4 \ll m_{Z^\prime}^4/2$, our results remain unchanged. See. Eq.~(\ref{eq:mass_phi}).}. 
Using Eq.~(\ref{RGEsol}), the quartic coupling is determined as a function of $\phi$, $\overline{\alpha_X}$, $m_{Z^\prime}$ and $x_H$. 
On the other hand, in the inflation analysis, the inflationary predictions are controlled by only one parameter $\xi$. 
Once we fix a $\xi$ value, $\phi_0$ and $\lambda_\Phi(\phi_0)$ are completely fixed as listed in Table~\ref{Tab:1}. 
Hence, by using Eq.~(\ref{RGEsol}) we can express $\overline{\alpha_X}$ as a function of $m_{Z^\prime}$ and $x_H$ 
  for a fixed value of $\xi$. 
In fact, for $\xi \lesssim 10$, we find that $\overline{\alpha_X}$ is almost independent of $x_H$, 
 and the $x_H$ dependence for inflationary predictions effectively drops off. 
Therefore, the inflationary predictions, $\overline{\alpha_X}$, $m_{Z^\prime}$, $m_\phi$ and $\theta$ 
  are directly related with each other through Eqs.~(\ref{eq:mass_phi}), (\ref{eq:theta}) and (\ref{RGEsol}).

\begin{figure}[htb!]
   \begin{center}
   \includegraphics[scale=0.7]{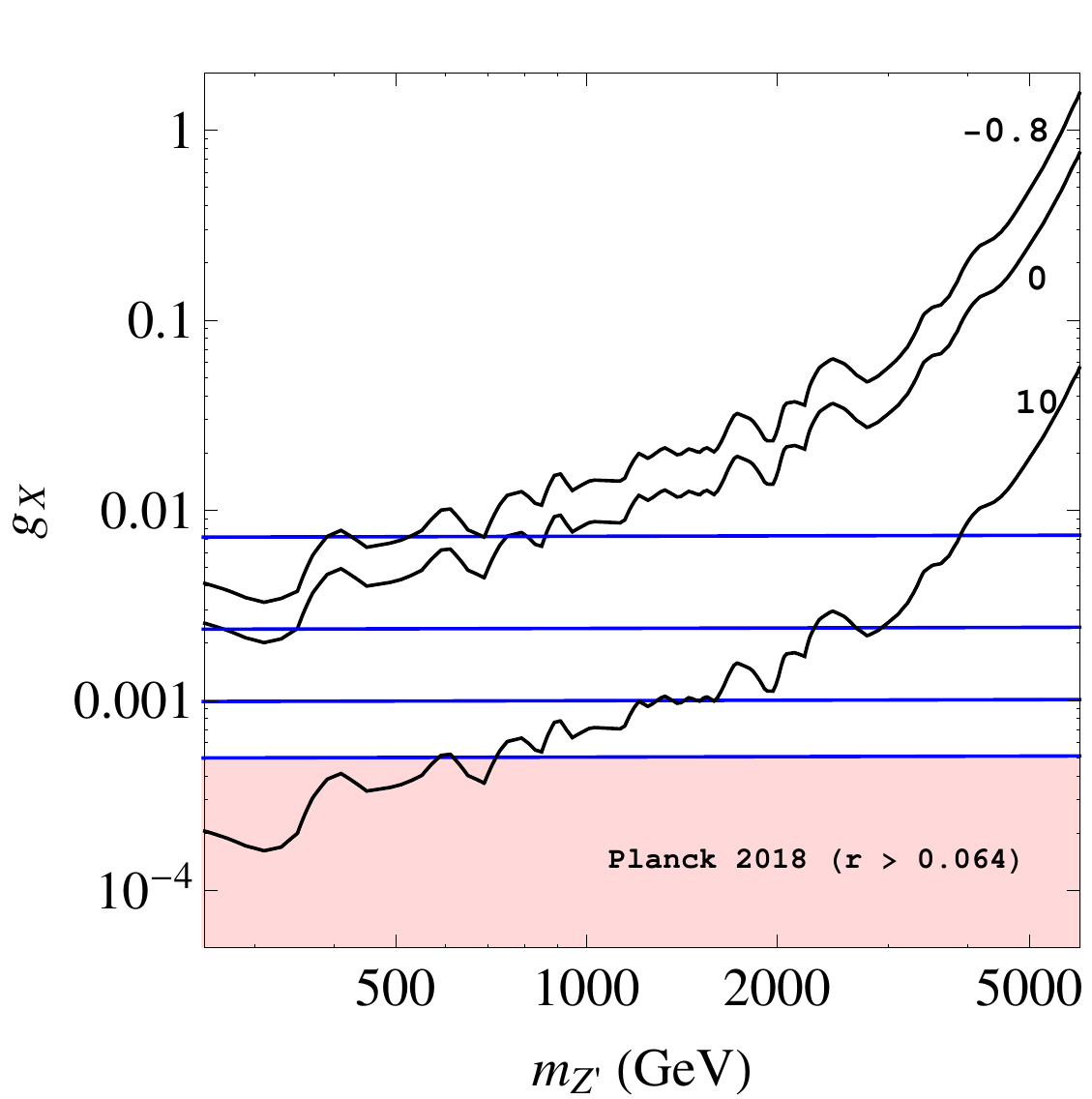} 
   \end{center}
\caption{
The upper bounds on ${\overline {g_X}}$ from the ATLAS result for $x_H=-0.8$, 0 and 10 (the diagonal black lines from top to bottom), respectively. 
For $x_H=0$, the horizontal blue lines from top to bottom correspond to 
   $\xi = 10$, 1.0, $0.12$, and $9.8\times 10^{-3}$ 
   or equivalently, $r = 4.3\times 10^{-3}, 4.9\times 10^{-3}, 0.01$,  and $0.064$, respectively. 
The red shaded region is excluded by the Planck 2018 measurements. 
}
\label{fig:fig1}
\end{figure}

The ATLAS and the CMS collaborations have been searching for  a narrow resonance at the LHC, 
  and the most severe constraint on the $Z^\prime$ boson of our model has been obtained  
  by the search with dilepton final states. 
The ATLAS collaboration has recently reported their final result of the LHC Run-2  
  with a 139 fb$^{-1}$ integrated luminosity \cite{Aad:2019fac}. 
Following the analysis in Ref.~\cite{Das:2019fee}, we interpret the ATLAS result  
  into an upper bound on $\overline{g_X}$ as a function of $m_{Z^\prime}$ for a fixed $x_H$ value. 
In Fig.~\ref{fig:fig1}, we show our results for $x_H = -0.8$, 0, and 10 (the diagonal black lines from top to bottom). 
The upper bounds depend on $x_H$ values and roughly scale as $\overline{g_X}/|x_H|$ for $|x_H| \gtrsim 3$, 
  while we find the LHC bound becomes weak for $x_H \sim -1$ \cite{Okada:2016tci}. 
In the figure, we also plot the contours for fixed $\xi$ values. 
For $x_H=0$, the horizontal blue lines from top to bottom correspond to 
   $\xi = 10$, 1.0, $0.12$, and $9.8\times 10^{-3}$ 
   or equivalently, $r = 4.3\times 10^{-3}, 4.9\times 10^{-3}, 0.01$,  and $0.064$, respectively. 
The red shaded region is excluded by the Planck 2018 measurement, $r > 0.064$. 
As discussed above, the inflationary predictions are almost independent of $x_H$ for $|x_H| < 10$
  and the horizontal lines represent the results for any values of $x_H$ for $|x_H| < 10$. 
Fig.~\ref{fig:fig1} indicates a complementarity between the LHC search for the $Z^\prime$ boson resonance 
  and the inflationary predictions. 
Since the $Z^\prime$ boson production cross section is dominated by the narrow resonance point, the cross section is proportional to $g_X^2$ in the narrow decay width approximation. Hence, we naively expect that the upper bound on $g_X^2$ to scale as luminosity, so that the future bound on $g_X \simeq g_X ({\rm current}) \times \sqrt{\frac{139/{\rm fb} } {{\cal L}/ {\rm fb}}}$.

\begin{figure}[t!]
   \begin{center}
   \includegraphics[scale=0.45]{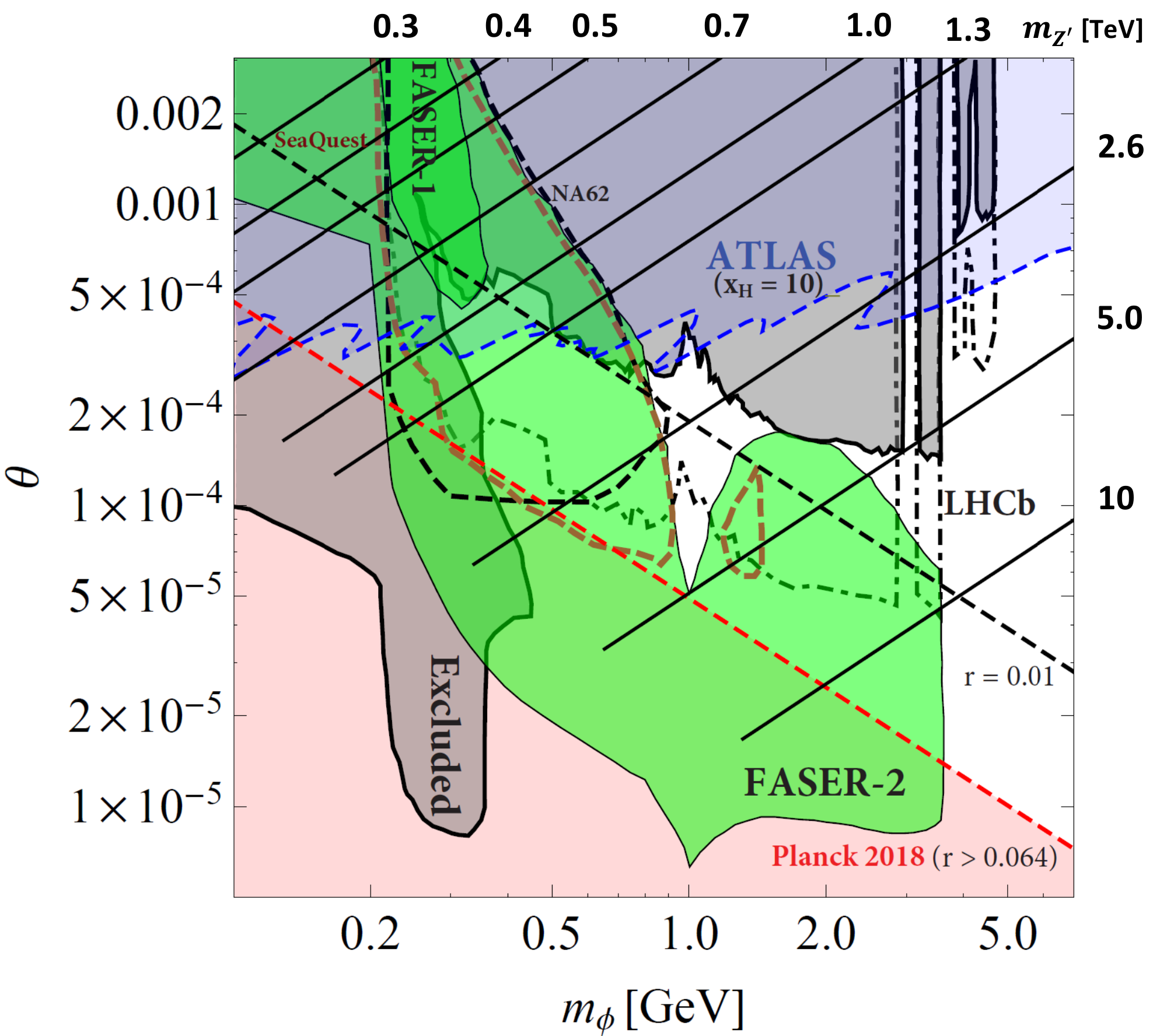}   
   \end{center}
\caption{
 The inflaton search reach at FASER (green shaded region)  and the relation with other observables. 
The diagonal dashed lines correspond to $\xi = 9.8 \times 10^{-3}$ ($r = 0.064$) and $\xi = 0.12$ ($r = 0.01$), respectively, from left to right. 
The diagonal solid lines correspond to $m_{Z^\prime} [{\rm TeV}] = 0.3$, 0.4, 0.5, 0.7, 1.0, 1.3, 2.6, 5.0, and 10, 
  from top to bottom.   
The blue shaded region (labeled ATLAS) is excluded by the ATLAS result of the $Z^\prime$ boson search for $x_H = 10$, corresponding to the bottom solid line in Fig.~\ref{fig:fig1}. 
The red shaded region is excluded by the Planck 2018 measurements.  
 }
\label{fig:fig2}
\end{figure}

{\bf Searching for the inflaton at FASER}: 
We are now ready to discuss the inflaton search at FASER 
  and its complementarity to the cosmological constraints on the inflationary predictions. 
For a fixed $\xi$ value, the inflationary predictions are fixed and  
  $\overline{\alpha_X}$ is determined as a function of $m_{Z^\prime}$, 
  independently of $x_H$  for $|x_H| < 10$.  
As a result, both the mass of inflaton ($m_\phi$) and its  mixing angle with the SM Higgs field ($\theta$) are uniquely determined by the CW relations in Eqs.~(\ref{eq:mass_phi}) and (\ref{eq:theta}), respectively.

In Fig.~\ref{fig:fig2}, we show our results\footnote{The search reach and current experimental bound presented in Fig.~2 applies to any SM singlet scalar field.} in ($m_\phi, \theta$)-plane, 
   together with FASER search reach, the search reach of 
   other planned/proposed experiments (contours with the names of experiments indicated), 
   and the current excluded region (gray shaded) from CHARM \cite{Bergsma:1985qz}\footnote{The updated analysis in Ref.~\cite{Winkler:2018qyg} shows that the bound on $\theta$ for $m_\phi \simeq 0.3$ GeV could be significantly relaxed. Also, see Refs.~\cite{Microboone} for recently updated bounds for $m_\phi = {\cal O} (0.1)$ GeV.}, Belle \cite{Wei:2009zv} and  LHCb \cite{Aaij:2015tna} experiments, as shown in Ref.~\cite{Ariga:2018uku}. 
Here, to ensure the readability of Fig.~\ref{fig:fig2}, we have not shown the search reach of other experiments such as SHiP \cite{Alekhin:2015byh}, MATHUSLA \cite{Evans:2017lvd} and CODEX-b \cite{Gligorov:2017nwh} presented in Ref.~\cite{Ariga:2018uku}. 
After our analysis, each point in FASER parameter
space has a one-to-one correspondence with inflationary predictions and $Z^\prime$ boson search parameters.
The diagonal dashed lines correspond to $\xi=9.8\times10^{-3}$ ($r=0.064$) and $\xi=0.12$ ($r=0.01$), 
  respectively, from left to right.   
The light red shaded region ($r > 0.064$) is excluded by the Planck 2018 results. 
We find that the parameter region corresponding to the inflationary prediction $r \gtrsim 0.01$ 
  can be searched by FASER 2 in the future, 
a part of which is already excluded 
  the Planck 2018 result. 
For a fixed $m_{Z^\prime}$, we can obtain a relation between $m_\phi$ and $\theta$ 
  through $\overline{\alpha_X}$ (recall, again, that this relation is almost independent of $x_H$ values for $|x_H|< 10$). 
In Fig.~\ref{fig:fig2}, the diagonal solid lines correspond to $m_{Z^\prime} [{\rm TeV}] = 0.3$, 0.4, 0.5, 0.7, 1.0, 1.3, 2.6, 5.0, and 10, 
  from top to bottom.   
A point on a solid line corresponds to a fixed value of $\xi$, or equivalently, $r$. 
Along each line, the $\xi$ ($r$) value increases (decreases) from left to right. 
For example, the left (right) diagonal dashed lines denote $r = 0.0064$ and $r = 0.01$. 
In Table~\ref{Tab:2}, for various $m_{Z^\prime}$ values, 
  we have listed the range of the predicted tensor-to-scalar ratio ($r$) 
  which will be covered by FASER.  
The blue shaded region (labeled ATLAS) is excluded by the ATLAS result of the $Z^\prime$ boson search 
   for $x_H = 10$, corresponding to the bottom solid line in Fig.~\ref{fig:fig1}. 
The excluded regions for $x_H = -0.8$ and $x_H =0$ (the $B-L$ model limit) correspond to $\theta > 10^{-3}$, 
   and thus they are covered by the gray shaded region.

\begin{table}[t!]
\begin{tabular}{|c|c|}
\hline
$m_{Z^\prime}$[TeV] &  The range covered by FASER \\ \hline
0.7    &    $4.2\times10^{-3}  \leq  r  \leq 9.0 \times 10^{-3}$   \\
1.0    &    $8.4\times10^{-3}  \leq  r  \leq  1.7\times 10^{-2}$  \\
1.3    &    $2.6\times10^{-3}  \leq  r  \leq  9.9 \times 10^{-3}$  \\
2.6    &    $1.3\times10^{-2}  \leq  r  \leq  6.4 \times10^{-2}$  \\
5.0    &    $7.3\times10^{-3}  \leq  r  \leq  6.4\times10^{-2}$ \\
10     &    $1.1\times10^{-2}  \leq  r  \leq  6.4 \times10^{-2}$ \\
\hline
\end{tabular}
\caption{ 
The ranges of $r$ which will be covered by FASER. 
} 
\label{Tab:2}
\end{table}

\begin{figure}[th!]
   \begin{center}
   \includegraphics[scale=0.65]{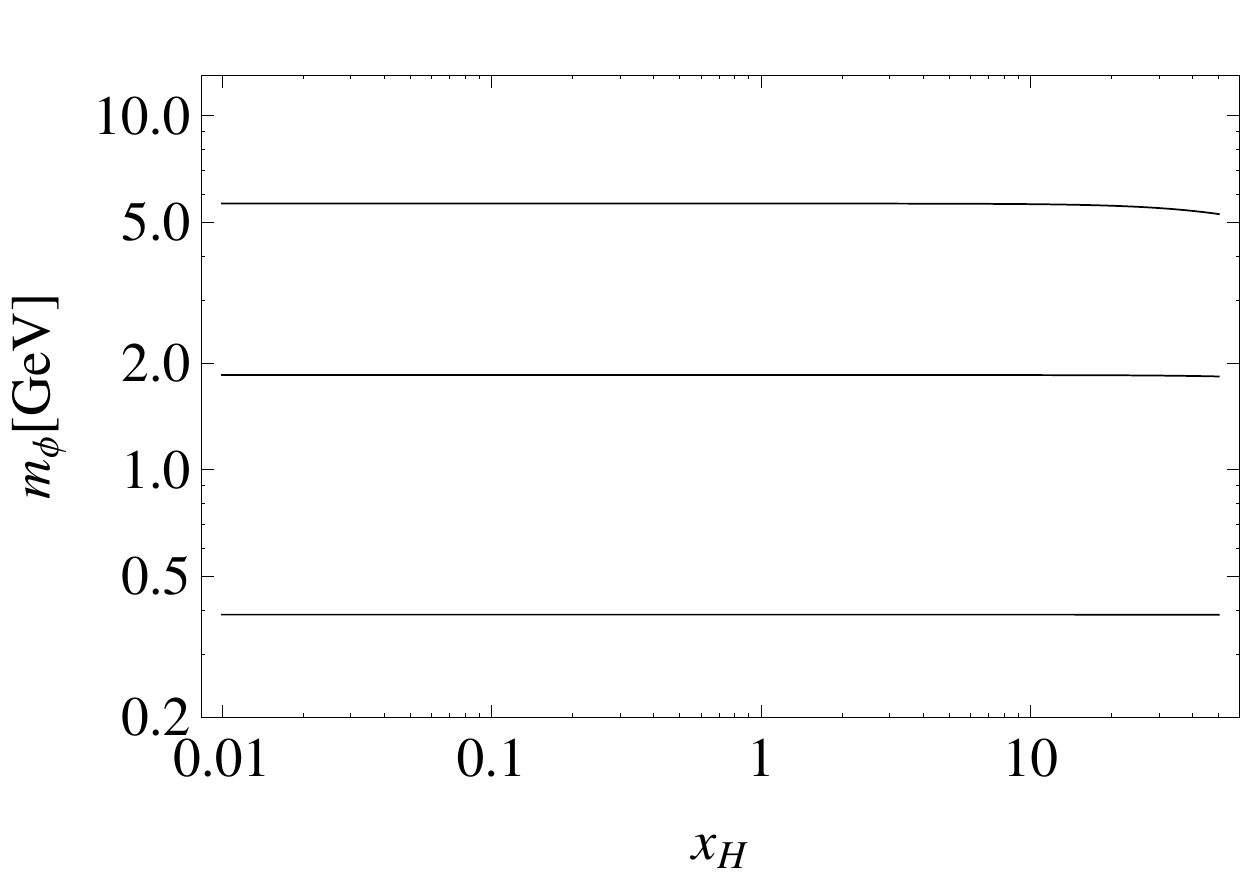} 
   \end{center}
\caption{
For fixed $m_{Z^\prime}= 2$ TeV, the horizontal solid lines from top to bottom correspond to 
   $\xi = 10, 1.0$, and $9.8\times 10^{-3}$ 
   or equivalently, $r = 4.3\times 10^{-3}, 4.9\times 10^{-3}$, and $0.064$, respectively. 
}
\label{fig:fig3}
\end{figure}
{\bf Inflaton decay and reheat temperature}: 
To complete our discussion of inflation scenario, lets us now  discuss reheating after inflation which proceeds via decay of the inflaton to SM particles. 
In our case, the inflaton decays into light SM fermions through the mixing with the SM Higgs boson. 
Using Eq.~(\ref{eq: massmatrix}), the decay width of the inflaton into SM particles can be expressed as
\bea
\Gamma({ \phi} \to SM) 
 &\simeq& \theta^2 \times \Gamma_ {{SM}} (m_{\phi}),
\label{eq:phitoSM}
\eea 
respectively, where $\Gamma_ {SM}(m_\phi)$ is the total decay width of the SM Higgs boson if its mass were $m_\phi$. 
Using the inflaton decay width, we estimate the reheat temperature by 
\bea 
  T_{\rm R} = \left( \frac{90}{\pi^2 g_*} \right)^{1/4} \sqrt{\Gamma({ \phi} \to SM) \times  M_P}, 
\label{eq:TR}
\eea  
where $g_* $ is the total effective degrees of freedom of thermal plasma. 
Therefore, the reheat temperature is also determined by the two parameters $m_\phi$ and $\theta$ used in the preceding analysis. 
Both $m_\phi$ and $\theta$ are uniquely determined by $\xi, x_H$ and $m_{Z^\prime}$ values. 
In Fig.~\ref{fig:fig3}, we show $m_\phi$ as a function of $x_H$ for fixed $m_{Z^\prime} = 2$ TeV. 
The horizontal solid lines from top to bottom correspond to 
   $\xi = 10, 1.0$, and $9.8\times 10^{-3}$ 
   or equivalently, $r = 4.3\times 10^{-3}, 4.9\times 10^{-3}$, and $0.064$, respectively. 
We see  that $m_\phi$ is independent of $x_H$ for $x_H \leq 10$

\begin{figure}[t!]
   \begin{center}
   \includegraphics[scale=0.9]{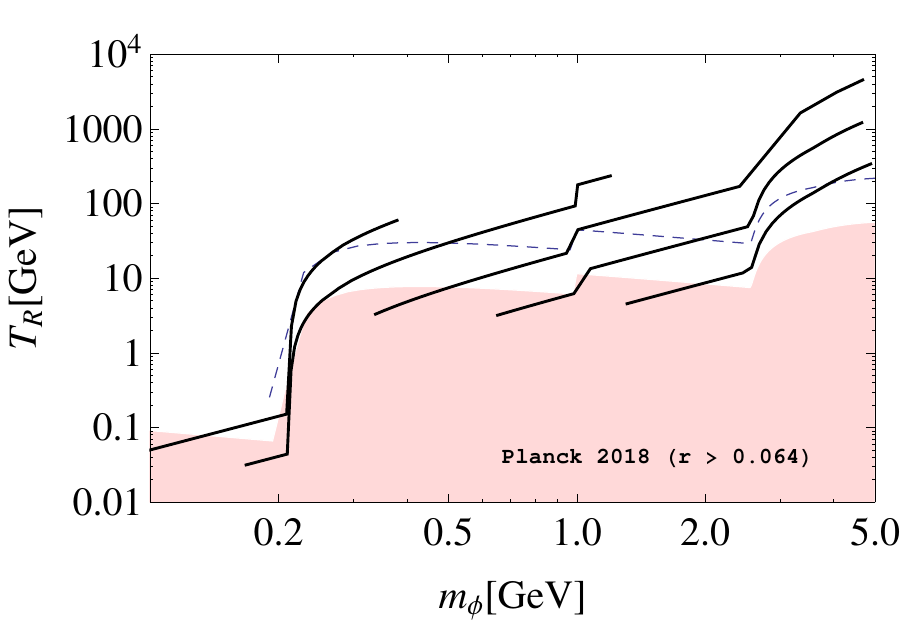} 
   \end{center}
\caption{
For $x_H = 0$, the solid lines from left to right correspond to $T_R$ contours for fixed $m_{Z^\prime} [{\rm TeV}] = 0.7, 1.3, 2.6, 5.0,$ and $10$, respectively. 
The red shaded region is excluded by the Planck 2018 measurements corresponding to $r>0.064$ while the dashed lines correspond to $\xi=0.12$ ($r=0.01$). 
}
\label{fig:fig4}
\end{figure}
In Fig.\ref {fig:fig4}, we show $T_R$ as a function of $m_\phi$ for fixed $x_H = 0$. 
The lines from left to right correspond to contours for fixed  $m_{Z^\prime} [{\rm TeV}] = 0.7, 1.3, 2.6, 5.0,$ and $10$, respectively. 
The red shaded region corresponds to $\xi=9.8\times10^{-3}$ ($r>0.064$) and is excluded by the Planck 2018 measurements. 
The dashed line corresponds to $\xi=0.12$ ($r=0.01$).  
For each contours in Fig.\ref{fig:fig4}, we have constrained the inflaton mass to lie in the range $0.1 < m_\phi {\rm [GeV]} \leq 5$ and the mixing angle satisfies $\theta < 10^{-3}$.  
For a fixed value of $m_\phi$, there is a one-to-one correspondence between the reheat temperature $T_R$ in Fig.\ref {fig:fig4} and the mixing angle $\theta$ in Fig.~\ref {fig:fig2}. 
Let us approximate the decay width 
\bea 
  \Gamma_ {{SM}} (m_{\phi}) \simeq  \frac{N_c}{8\pi} \left(\frac{m_f}{v_h}\right)^2 m_\phi, 
\eea  
where $N_c$ is the color factor for the fermion and anti-fermion in the final state with a mass $m_f < m_\phi /2$. 
Together with Eqs.~(\ref{eq:phitoSM}) and (\ref{eq:TR}), we obtain  
\bea
\theta \simeq \frac{8.23\times10^{-5}}{{\sqrt N_c}} \left(\frac{0.1}{m_f  }\right)  \left(\frac{T_R}{5}\right)   \left(\frac{1}{\sqrt{m_\phi}} \right), 
\eea 
where $m_f$, $T_R$, and $m_\phi$ are in GeV units.

In our analysis we have considered the number of e-folds $N_0$ as a free parameter and fixed $N_0 = 50$. 
However, $N_0$ is determined by $T_R$ and tensor-to-scalar ratio $r$ (or, equivalently the Hubble during the inflation) as 
\bea
N_0 \simeq 49.21 + \frac{1}{6} \left(\frac{r}{0.01}\right)+ \frac{1}{3} \left(\frac{T_R}{100 \;{\rm GeV}}\right)
\eea 
This is consistent with our choice $N_0 = 50$. 
We have also checked that the inflationary prediction for a fixed $\xi$ value weakly depends on $N_0$ values.

In conclusion, we have considered the non-minimal quartic inflation scenario 
  in the minimal U(1)$_X$ model with classical conformal invariance, 
  where the inflaton is identified with the U(1)$_X$ Higgs field.  
FASER can search for the inflaton when its mass and mixing angle with the SM Higgs field 
  are in the range of $0.1 \lesssim m_\phi[{\rm GeV}] \lesssim 4$ and $10^{-5} \lesssim \theta \lesssim 10^{-3}$. 
By virtue of the classical conformal invariance and the radiative U(1)$_X$ symmetry breaking
   via the Coleman-Weinberg mechanism, 
   the inflaton search by FASER, the $Z^\prime$ boson resonance search at the LHC, 
   and the future measurement of $r$ are complementary to test our inflationary scenario.

{\bf Acknowledgements:} 
This work is supported in part by the United States Department of Energy grant DE-S0012447 (N.O), DE-SC0013880 (D.R)
  and Bartol Research Grant BART-462114 (D.R). 


\end{document}